\documentclass[prl,priprint,epsf,psfig]{revtex4}
\usepackage{graphicx}
\DeclareGraphicsExtensions{.pdf,.png,.gif,.jpg}
\usepackage{float}
\usepackage{subfig}
\usepackage{epsfig}
\usepackage{wrapfig}

\begin{document}

\title{Phase diagram of strongly attractive $p$-orbital fermions on optical lattices}

\author{Theja N. De Silva}
\affiliation{Department of Chemistry and Physics,
Georgia Regents University, Augusta, GA 30912, USA.}
\begin{abstract}
We examine a system of doubly degenerate $p$-orbital polarized fermions on a two-dimensional square lattice with a strong on-site interaction. We consider the system density at the half filling limit and tackle the strong attractive interaction using a perturbation theory. We treat the four-site square plaquette interaction term generated from the directional tunneling dependence of $p$-orbitals using the fourth order in perturbation theory. We map the strong coupling particle Hamiltonian into an effective spin-Hamiltonian and then use a variational mean field approach and a linear spin-wave theory to study the phase diagram. Further, we discuss the experimental signatures of these phases within the context of current cold-atom experimental techniques.
\end{abstract}

\maketitle

\section{I. Introduction}

Cold atoms loaded into an optical lattice formed by interference of counter propagating laser beams are regarded as a quantum simulator for many-body condensed matter systems~\cite{intro}. The flexibility of generating various lattice geometries and the controllability of the dimension, lattice parameters, and interaction parameters within the cold atomic setups make them ideal test beds for studying strongly correlated many-body phenomena. In condensed matter many-body systems, such as transition metal oxides and rare-earth materials, both spin and charge degrees of freedom play an important role determining their properties. For some materials, the orbital degrees of freedom is also an vital factor. The orbital degrees of freedom is active when the electronic orbitals are degenerate and partially filled with electrons. The strongly correlated nature and spatial anisotropy of $d$ and $f$ orbital electrons in some correlated materials are responsible for various exotic magnetic, superconducting, and transport properties.

Due to the recent advancement of laser technology and the progress in experimental techniques, cold atomic physics and condensed matter physics are in close proximity to each other. Condensed matter models can be engineered using atoms trapped in a combined harmonic and lattice potentials using various laser arrangements and other experimental laser techniques in cold atom setups. These so-called quantum simulators or atoms in optical lattices, not only provide better opportunities to explore existing condensed matter many-body phenomena, but also allow one to seek and explore new states of matter and their physical properties. Unlike solid state electronic models, the optical lattice setups are rigid and robust. As a result, the Jahn-Teller distortion and the conventional phonons are absent.  However, these effects can be engineered using various laser techniques and different types of atoms at low temperatures. The electrons which are fermions are the constitute particles in condensed matter lattice models. However, not only fermions but also bosons, as well as mixture of fermions and bosons can be used as constitute particles in optical lattice models. Opposed to the condensed matter systems, even $p$-orbital occupation of atoms in optical lattices can show strongly correlated many-body phenomena. Recently, a tremendous experimental and theoretical progress has been achieved in the direction of understanding $p$-orbital bands of both fermions and bosons in optical lattices. Fermions on a $p$-orbital band have been imaged and probed by transferring them into the $p$-orbital band using a sweep across the Feshbach resonance~\cite{kohl}. As we consider in this letter, $p$-orbital bands can be activated even simpler way by completely filling the lowest $s$-orbital band with two pseudo-spin $1/2$ atoms and then having more than two atoms per site. The population of $p$-orbital band bosons were observed in moving lattice experiments~\cite{brow} and dynamically deformed double-well lattices~\cite{nist}. The $p$-orbital boson decay time has also been measured by pumping atoms by Raman transitions~\cite{meul}. The quantum mechanical phenomenon associates with orbital degrees of freedom of atoms in optical lattices can be captured by a $p$-orbital band Hubbard Hamiltonian~\cite{isac, vinc, maci}. The emergence of various quantum phases of $p$-orbital band bosons in optical lattices has been recently proposed~\cite{wu1, kukl, wu2, xu, hebe, boya, yong, xiao1, toma, zi, mcin, jani, stas, jona, mart, xiao2, huan, coll, umuc}. A tremendous progress in studying orbital exchange driven magnetic properties of $p$-orbital band fermions has also been reported~\cite{xian, wu3, wang, zhang, zhao, wu4, xiao3, yaji, wino, zixu1, giaw, hsia, phil, mach, zixu2, wu5}.

In this letter, we study the phase diagram of $p$-orbital attractive polarized fermions loaded into a optical lattice. The $p$-orbital occupation can be achieved by having two one-component fermions at each site. In this scenario, the $s$-orbital is completely filled and the $s$-orbital atom remains inert at the low energy subspace. The attractive interaction can be generated simply by adjusting the two-body scattering length to be negative values using Feshbach resonance. We consider the system density at the half filling and strongly interacting limits. At the half filling density limit, on average, \emph{only} one atom resides in a $p$-orbital at each site. First, we map the system Hamiltonian into an effective spin Hamiltonian. The fourth order tunneling process favored by the spatial anisotropy of the $p$-orbital orientation is taken into account using the fourth order in our perturbation theory. Second, we map out the magnetic phase diagram of the effective spin Hamiltonian using a variational mean field theory. Third, we use a linear spin-wave theory to study the phase diagram and magnon excitation spectrum.

The letter is organized as follows. In section II, we introduce the effective model and discuss the details of mapping it into an effective spin-model. In section III, we discuss our variational mean field approach and provide the resulting phase diagram of the effective spin-model. In section IV, we provide the details of our linear spin-wave theory for the effective spin-model and discuss the phase diagram and magnon excitation spectrum. In section V, we discuss the connection of the resulting phase diagram to the original model and experimental signatures of the magnetic phases. Finally in section V, we draw our conclusions.

\section{II. Effective Model}

\begin{figure}
\includegraphics[width=0.65 \columnwidth]{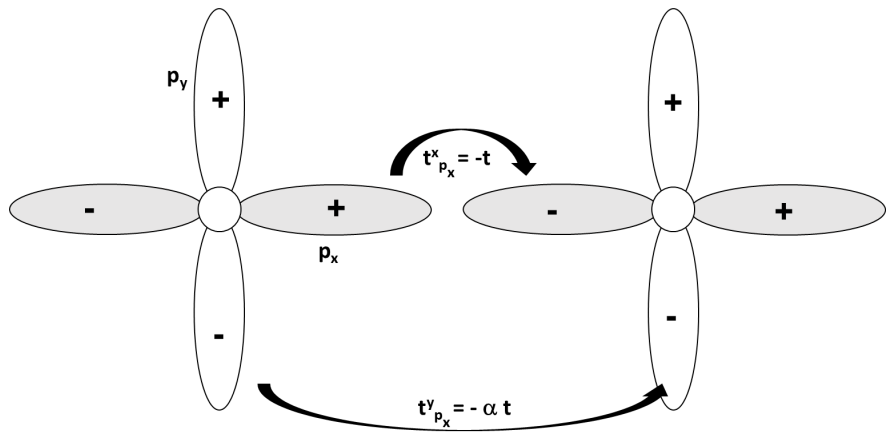}
\caption{While the tunneling amplitude $t$ along the $x$-direction between the $p_x$ orbitals is large, the tunneling amplitude $\alpha t$, (with $\alpha \ll 1$) between the $p_y$ orbitals is much weaker. The gray region represents the $p_x$ orbital, whereas the white region represents the $p_y$ orbital.}\label{h2s}
\end{figure}

We consider multi-orbital system of polarized fermions on a two-dimensional square lattice. We assume that the atoms are loaded to the lattice such that the $s$-orbital is completely filled and the dynamic of the system is determined only by the $p$-orbital atoms. Further, we assume that only $p_x$ and $p_y$ orbitals are active on the two-dimensional (2D) lattice. This scenario is reasonable as the $s$ and $p$-orbital bands are separated by a large band gap at experimentally relevant parameters and $p_z$ orbital has a higher energy in 2D configurations. By imposing a strong laser beam along the $z$-direction, the $p_z$-orbital band energy can further be increased. As sketched in FIG.~\ref{h2s}, the $p_x$ orbitals overlap only in the $x$-direction. As a result, the tunneling between $p_x$-orbitals in $x$-direction is much greater than that of the $y$-direction. Similarly, the hopping between $p_y$ orbitals along $y$-direction is greater than that of the $x$-direction. By representing the $p_x$ orbital occupation by pseudo-spin $\uparrow$ and $p_y$ orbital occupation by pseudo-spin $\downarrow$, the dynamics of the spin polarized atoms in pseudo-spin-dependent 2D optical lattice can be described by an effective Hamiltonian,

\begin{eqnarray}
H = - \sum_{<ij>,\sigma} t^{ij}_\sigma c^\dagger_{i\sigma}c_{j\sigma}- U\sum_i(n_{i\uparrow}-1/2) (n_{i\downarrow}-1/2) - \mu \sum_{i\sigma} c^\dagger_{i\sigma}c_{i\sigma}+ V\sum_{<ij>}n_i n_j \label{H1}.
\end{eqnarray}

\noindent The first term is the kinetic energy and is proportional to the tunneling amplitude $t^{ij}_\sigma$ along the bond $ij =x, y$ directions of $\sigma = \uparrow (p_x),  \downarrow (p_y)$ atoms. The operator $c^\dagger_{i\sigma} (c_{i\sigma})$ creates(destroys) a Fermi atom with pseudo-spin  $\sigma = \uparrow, \downarrow$ at lattice site $i$. The second term describes the on-site interaction energy $U >0$. The density operator or the occupation number operator is $n_{i \sigma}= c^\dagger_{i\sigma} c_{i\sigma}$ and $\mu$ is the chemical potential. Notice $<ij>$ indicates only the nearest neighbor pair of sites and we neglect tunneling beyond the nearest neighbors. The last term represents the nearest-neighbor weak, off-site repulsive interaction. All other nearest neighbor interactions, such as bond-charge, spin-spin, and pair-hopping are neglected.

We consider the system at the half filling density, so that on average, only one atom resides at each lattice site occupying either $p_x$ or $p_y$ orbital. Due to the directional dependence of $p$-orbitals, we take the tunneling amplitude of atoms $t^{ij}_\sigma$ as $t^{ij = x}_\uparrow = t^{ij = y}_\downarrow = t$ and $t^{ij = x}_\downarrow = t^{ij = y}_\uparrow = \alpha t$ with $\alpha \ll 1$. As shown by the Hamiltonian above, we assume on-site interaction is attractive ($-U < 0$), which can be achieved by tuning the free-space two-body scattering length to be negative values. The directional pseudo-spin tunneling amplitudes and the on-site attractive interaction favor a four-site ring-exchange or plaquette interaction term in the form,

\begin{eqnarray}
H_P = K \sum_{<ijkl>,\sigma} c^\dagger_{i\sigma}c_{j\sigma^\prime} c^\dagger_{k\sigma^\prime }c_{l\sigma} \label{Hp},
\end{eqnarray}

\noindent where $<ijkl>$ represents the four-site square plaquette and the four spin coupling constant $K \propto t^4/U^3$ can be derived in fourth order hopping process. This process is schematically shown in FIG.~\ref{h4s}.

\subsection{Psudo-spin Model at the strong coupling Limit}

In the strongly interacting limit where $t \ll V < U$, the tunneling term acts as a small perturbation. In this section, we derive an effective pseudo-spin Hamiltonian in fourth order perturbation theory. The local Hilbert space has four states, namely empty site denoted by $|e \rangle$, singly occupied sites with up $|\uparrow \rangle$ or down $|\downarrow \rangle$ spins, and a doubly occupied site with opposite pseudo-spins $|d \rangle$. The energies of these states in unperturbed Hamiltonian, i.e the on-site part of the Hamiltonian, are given by $E_i = -U/4$, $E_i = -\mu + U/4$, $E_i = -\mu + U/4$, and $E_i = -2\mu - U/4$, respectively. Note that our notations up and down spins represent the $p_x$ and $p_y$ orbital occupation. At the half filling density limit where $\mu = 0$, the singly occupied sites have higher energies and the empty and double occupied sites energies are degenerate. The local Hilbert space of two-neighboring sites has ten states, however six of those states that involve a single site have higher energies. Therefore, the low-energy subspace constitutes only the four states, $|e, e \rangle$, $|e, d\rangle$, $|d, e \rangle$, and $|d, d \rangle$ with energies $E = -U/2$, $E = -2\mu - U/2$, $E = -2\mu - U/2$, and $E = -4\mu - U/2 +4V$, respectively. Here $|e, d\rangle$ stands for two neighboring sites with a empty and a doubly occupied sites and other notations have the similar meaning.

\begin{figure}
\includegraphics[width=0.8\columnwidth]{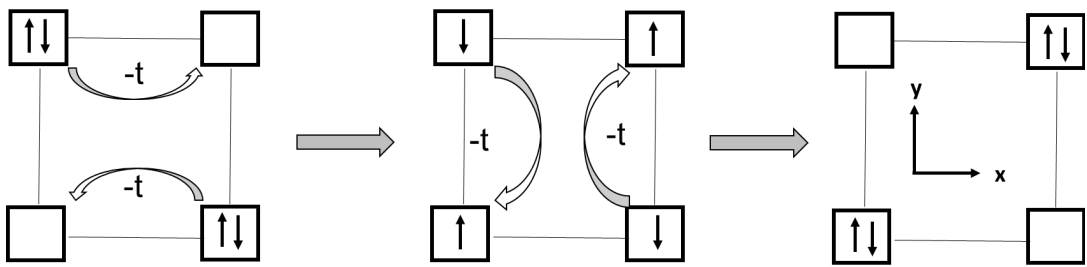}
\caption{Four site plaquette interaction at the half filling density limit for strongly attractive fermions due to the spatial anisotropy of $p$ orbitals. While the up-arrow ($\uparrow$) represents the $p_x$ orbital occupation, the down-arrow ($\downarrow$) represents $p_y$ orbital occupation. }\label{h4s}
\end{figure}

In the second order perturbation theory, the energy correction due to the tunneling between two neighboring sites can be represented as a matrix $M$,

\begin{widetext}
\begin{eqnarray}
M = \left(
      \begin{array}{c}
        \langle e, e| \\
        \langle e, d|  \\
        \langle d, e|  \\
        \langle d, d|  \\
      \end{array}
    \right)^\emph{T} \left(
                \begin{array}{cccc}
                  -U/2 & 0 & 0 & 0 \\
                  0 & -2\mu-U/2-(\alpha^2+1)t^2/(U+V) & -2\alpha t^2/(U+V) & 0 \\
                  0 & -2\alpha t^2/(U+V) & -2\mu-U/2-(\alpha^2+1)t^2/(U+V) & 0 \\
                  0 & 0 & 0 & -4\mu-U/2 \\
                \end{array}    \right) \left(
                              \begin{array}{c}
                                | e, e\rangle \\
                                | e, d\rangle  \\
                                | d, e\rangle  \\
                                | d, d\rangle  \\
                              \end{array}
                        \right).
\end{eqnarray}
\end{widetext}

Here we neglect the fourth order corrections as the dominant fourth order correction is taken into account through Hamiltonian $H_P$. The fourth order correction here is much smaller as it is proportional to $\alpha^2$ or its higher powers. The $\emph{T}$ in the column vector in Eq. (3) indicates the transpose. At the half filling density limit, we map this low-energy Hamiltonian into an effective spin-1/2 system by mapping the local double occupied sites by pseodo-spin-ups and the local empty sites by pseodo-spin-downs, ie $|d \rangle \rightarrow |\uparrow \rangle_{spin}$ and $|e \rangle \rightarrow |\downarrow \rangle_{spin}$, such that the z-component of the spin operators $S_z = \hbar \sigma^z/2$ acts as $\sigma^z |\uparrow \rangle_{spin} = +|\uparrow \rangle_{spin}$ and $\sigma^z |\downarrow \rangle_{spin} = -|\downarrow \rangle_{spin}$. Notice that these new pseodo-spin states $|\uparrow \rangle_{spin}$ and $|\downarrow \rangle_{spin}$ are different from our previous pseodo-spin states $|\uparrow \rangle$ or $|\downarrow \rangle$ that represented the $p$-orbital occupations. We start with the most general two site effective Hamiltonian $H_{ij}$ that respects the global SU(2) symmetry. Up to the quadratic terms, the spin-1/2 effective Hamiltonian must have the form $H_{ij} = \sum_\nu \{A_\nu S_i^\nu S_j^\nu + B_\nu S_i^\nu\} +C$, where $\nu = x, y, z$. By calculating the matrix elements with respect to this effective Hamiltonian $H_{ij}$ and equating them to the matrix elements in Eq. (3), all exchange interactions terms, $A_\nu$, $B_\nu$, and $C$ are derived. In terms of Pauli matrices $\vec{\sigma}$ that represent the new pseodo-spins, the effective pseodo-spin Hamiltonian is then written as,

\begin{eqnarray}
H = \sum_{<ij>} \{-J (\sigma^x_i \sigma^x_j + \sigma^y_i \sigma^y_j)+\Delta \sigma^z_i \sigma^z_j\} + \frac{K}{4}\sum_{<ijkl>} \sigma^z_i \sigma^z_j \sigma^z_k \sigma^z_l\label{HS},
\end{eqnarray}

\noindent where $J = \alpha t^2/(U+V)$, $\Delta = (\alpha^2+1)t^2/[2(U+V)]$ and $K = 12 t^4 U/[(12V+3U)^3(8V+3U)]$. The last term is originated from the fourth order tunneling process shown in FIG.~\ref{h4s}. All other fourth order tunneling processes are much weaker for the present system. We have already set the half filling condition $\mu =0$ and ignored the constant energy shift of $C = -U/2-(\alpha^2+1)t^2/[2(U+V)]$. At half filling all the linear terms $B_\nu$ vanish.

Without the last quartic term, Eq. (4) represents the well-known ferromagnetic XXZ model. As the model contains both ferromagnetic exchange $J$ and antiferromagnetic exchange $\Delta$, the spin Hamiltonian is frustrated even on a geometrically non-frustrated two-dimensional square lattice with only nearest neighbor exchange interactions. This model can show finite temperature Berezinskii-Kosterlitz-Thouless transition without finite magnetic order and a Ising type phase with long range order~\cite{cuc}. The XXZ spin model is equivalent to the hard-core boson Hamiltonian with nearest neighbor hopping $t = 2J$ and nearest neighbor repulsive interaction $V= \Delta$~\cite{koh}. In bosonic language, the magnetic order in the XY plane corresponds to superfluid order, while the magnetic order in the Z direction corresponds to density order.

\section{III. A Variational Mean Field theory for the spin Hamiltonian}

In this section, we use a variational mean field approach to study the phase diagram originating from the competition between three terms in the spin Hamiltonian given in Eq. (4). Notice all coupling constants, $J$, $\Delta$, and $K$ are positive. While $J$ term favors a ferromagnetic spin ordering in $x-y$ plane in spin space, $\Delta$ term prefers an antiferromagnetic ordering in $z$-direction. However, the $K$ term discourages both of these ordering.

First, we break up the original square lattice into two interpenetrating square sublattices $A$ and $B$. Next, we take our \emph{normalized variational density} matrices for sublattice $A$ and $B$ as

\begin{eqnarray}
\rho_{A,B} = \frac{1}{2} + \frac{1}{2}m(\sin \theta \sigma^x \pm \cos \theta \sigma^z),
\end{eqnarray}

\noindent where $m$ and $\theta$ are variational parameters and the upper sign is for the sublattice $A$ and the lower sign is for the sublattice $B$. This choice gives us the sublattice magnetization $m_{A,B} = Tr(\rho_{A,B} \vec{\sigma}) \equiv \pm m \cos \theta \hat{z} + m \sin \theta \hat{x}$, here $Tr(L)$ represents a trace of a $2 \times 2$ matrix $L$. Thus, $m = 0$ represents the paramagnetic state and $m \neq 0$ represents three magnetically ordered states depending on the value of $\theta$. For $\theta = 0$, the system is an antiferromagnetic state with $\vec{m}_A = -\vec{m}_B$ and its staggered magnetization is in the $\vec{z}$-direction. For $\theta = \pi/2$, the system is a ferromagnetic state with $\vec{m}_A = \vec{m}_B$ and its magnetization lying along the $x$-direction. On the other hand when $\theta \neq 0$, the system is a canted antiferromagnet with $\vec{m}_A \neq \vec{m}_B$.

The free energy $F = E-k_BTS$ with energy $E$ and entropy $S$, can easily be calculated using $E = Tr(\rho H)$ and $S = -Tr(\rho \ln \rho)$, where $k_B$ is the Boltzman constant and $T$ is the temperature. Defining dimensionless free energy per site $f(m, \theta) = F/[\tilde{J} N]$ with energy scale $\tilde{J} = t^2/[U+V]$ and $N$ being the number of lattice sites in the two-dimensional square lattice, we find

\begin{eqnarray}
f(m, \theta) = -2\alpha m^2 - (\alpha-1)^2 m^2 \cos^2 \theta  + \frac{\kappa}{4}m^4 \cos^4 \theta  + \tau \biggr[\frac{1+m}{2}\ln\biggr(\frac{1+m}{2}\biggr) + \frac{1-m}{2}\ln\biggr(\frac{1-m}{2}\biggr)\biggr].
\end{eqnarray}

\noindent Here the dimensionless temperature $\tau = k_BT/\tilde{J}$ and the dimensionless four-site square plaquette coupling $\kappa = K/\tilde{J}$. The minimization of the free energy with respect to two variational parameters, $\partial f/\partial m =0$ and $\partial f/\partial \theta =0$ leads to two coupled mean field equations,

\begin{eqnarray}
-4\alpha m -2(\alpha-1)^2 m \cos^2 \theta + \kappa m^3 \cos^4 \theta + \frac{\tau}{2} \ln \biggr(\frac{1+m}{1-m}\biggr) = 0
\end{eqnarray}

\noindent and

\begin{eqnarray}
-m^2 \sin \theta  \cos \theta [\kappa m^2 \cos^2 \theta-2(\alpha-1)^2] =0.
\end{eqnarray}

\noindent By simultaneously solving these mean field equations, we find

\begin{eqnarray}
\cos^2 \theta = \left\{
                  \begin{array}{ll}
                    0, & \hbox{if $(\alpha-1)^2 < 0$;} \\
                    \frac{2(\alpha-1)^2}{\kappa m^2}, & \hbox{if $0 \leq (\alpha-1)^2 \leq \kappa m^2/2$;} \\
                    1, & \hbox{if $(\alpha-1)^2 > \kappa m^2/2$.}
                  \end{array}
                \right.\label{te}
\end{eqnarray}

\noindent Obviously, the first condition cannot be fulfilled, thus a $xy$-ferromagnetic phase is not stable for our effective spin-Hamiltonian in Eq. (4). Using the $theta$ values in Eq. (\ref{te}) and the first mean field equation, we find a canted antiferromagnetic (C-AFM) state for $\kappa > 2(1-\alpha)^2$ and $\tau < \tau_c$ where the critical dimensionless temperature is given by,

\begin{eqnarray}
\tau_c = \frac{4 \alpha \sqrt{2(1-\alpha)^2/\kappa}}{tanh^{-1}\sqrt{2(1-\alpha)^2/\kappa}}.
\end{eqnarray}

\begin{figure*}
\centering
\subfloat[The phase diagram at a fixed dimensionless temperature $\tau =1$.]{\includegraphics[width=0.5\columnwidth]{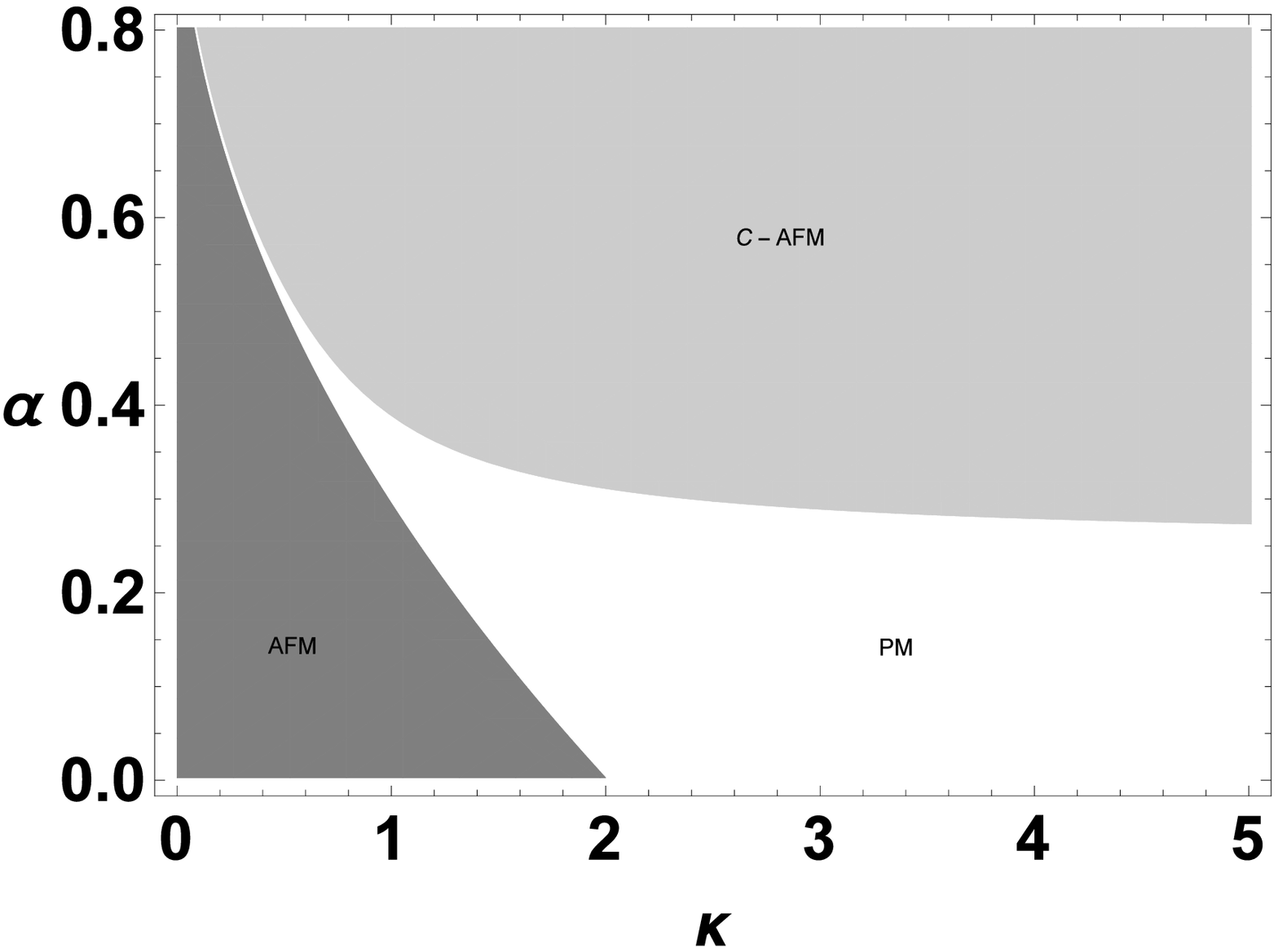}}
\subfloat[The phase diagram at a fixed tunneling anisotropy $\alpha =0.3$]{\includegraphics[width=0.5\columnwidth]{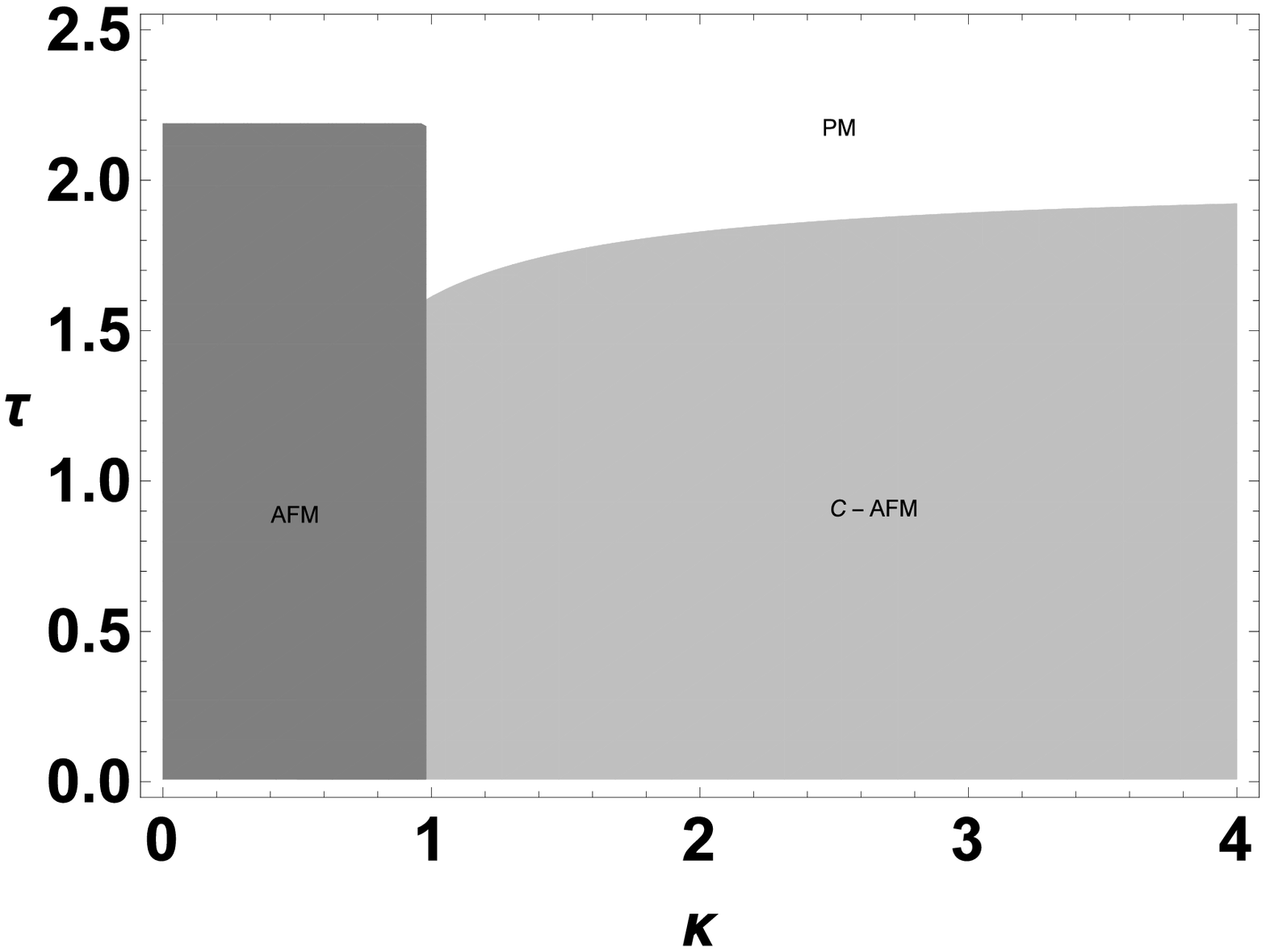}}
\caption{Magnetic phase diagram obtained from the variational mean field theory of Hamiltonian (4). Theory predicts Antiferromagnetic (AFM), canted antiferromagnetic (C-AFM), and paramagnetic (PM) phases in $\alpha-\kappa$ $\tau-\kappa$ planes, where $\alpha$ is the tunneling anisotropy and $\kappa$ is the the dimensionless four-site square exchange interaction parameter.} \label{pd}
\end{figure*}

\noindent On the other hand, for $\kappa < 2(1-\alpha)^2$, we find an antiferromagnetic (AFM) state if $\tau < 2(\alpha^2+1)$. If any of these conditions are not satisfied, the system remains in a paramagnetic (PM) state. The phase diagrams resulting from these conditions for various parameter regions are shown in FIG.~\ref{pd}.

\section{IV. A Linear spin-wave theory for the spin Hamiltonian}

In this section, we investigate the quantum phase transition between AFM state and C-AFM state using a linear spin-wave theory (LSWT). In quantum spin systems, the low energy or long-wave length magnetic excitations spectrum is formed by spin-waves or magnons. Magnons are collective modes associated with the coherent precession of spins at a given site. The original introduction of spin-wave concept was proposed by Bloch and Slater~\cite{lswt}. Holstein and Primakoff later introduced the quantization of spin waves in terms of bosonic operators that represent magnons~\cite{hp}. LSWT is the linearized version of this bosonic theory and it is a well known tool in the study of quantum spin systems. In this method, each spin operator is represented by two Holstein-Primakoff (HP) bosons~\cite{hpA}. The vacuum state of Holstein-Primakoff bosons will be a broken symmetric state, therefore the theory is valid \emph{only} for long-range magnetically ordered states. Some generalizations have been proposed to validate LSWT for short-range magnetically disordered states~\cite{mswt1, mswt2, mswt3, mswt4}. Here we assume that the spin system is in either AFM ordered state or C-AFM ordered state and study the phase transition between them. In LSWT, the quantum spin Hamiltonian is mapped onto an interacting boson Hamiltonian and then keep only the quadratic terms by neglecting the higher order bosonic terms. The higher order bosonic terms represent the interaction between HP bosons or magnons. Consequently, LSWT is a method of converting an interacting spin Hamiltonian into a non-interacting bosonic Hamiltonian.

First, we rotate the spin operators $\vec{\sigma}_i^{0}$ into the local reference frame at each site using Euler rotation $\vec{\sigma}_i = \underline{U}_i \vec{\sigma}_i^{0}$ [here $\vec{\sigma}_i^{0}$'s are the original spin operators in Eq. (4) and $\vec{\sigma}_i$'s are the spin operators in local reference frame], where the rotation matrix $\underline{U}_i$, is given by~\cite{cst1, cst2},

\begin{eqnarray}
\underline{U}_i = \left(
                    \begin{array}{ccc}
                      \cos \theta_i & 0 & -\sin \theta_i \\
                      0 & 1 & 0 \\
                      \sin \theta_i & 0 & \cos \theta_i \\
                    \end{array}
                  \right).
\end{eqnarray}

\noindent Notice that we consider only the in-plane rotation so that we have set the second Euler angle $\psi$ to be zero. In the local frame, the Hamiltonian in Eq. (4) reads $H = \sum_{\langle i\in A, j\in B \rangle} H_{AB} + H_q$,

\begin{eqnarray}
H_{AB} = J_x \sigma_{Ax} \sigma_{Bx} + J_y \sigma_{Ay} \sigma_{By} + J_z \sigma_{Az} \sigma_{Bz} + J_{zx} \sigma_{Az} \sigma_{Bx} + J_{xz} \sigma_{Ax} \sigma_{Bz},
\end{eqnarray}

\noindent where the four spin interaction terms $H_q = J_q \sum_{ijkl} \sigma_{iz}^0 \sigma_{jz}^0 \sigma_{kz}^0 \sigma_{lz}^0$ with $J_q = K/4$, in the local frame is given by,

\begin{eqnarray}
H_{q} = J_q\sum_{ i\in A, j\in B, k\in A, l\in B} \biggr\{ \cos^2 \theta_A \cos^2 \theta_B\sigma_{iz} \sigma_{jz} \sigma_{kz}\sigma_{lz} - \frac{1}{2} \sin (2\theta_A) \cos^2 \theta_B [\sigma_{iz}\sigma_{jz}\sigma_{kx}\sigma_{lz} + \sigma_{ix}\sigma_{jz}\sigma_{kz}\sigma_{lz}] \\ \nonumber + \sin^2\theta_A \cos^2\theta_B \sigma_{ix}\sigma_{jz} \sigma_{kx}\sigma_{lz} - \frac{1}{2}\cos^2 \theta_A \sin (2\theta_B) [\sigma_{iz} \sigma_{jz}\sigma_{kz}\sigma_{lx} + \sigma_{iz}\sigma_{jx}\sigma_{kz}\sigma_{lz}] \\ \nonumber + \frac{1}{4} \sin (2\theta_A) \sin (2\theta_B) [\sigma_{iz} \sigma_{jz}\sigma_{kx}\sigma_{lx} + \sigma_{ix}\sigma_{jz}\sigma_{kz}\sigma_{lx} + \sigma_{iz}\sigma_{jx}\sigma_{kx}\sigma_{lz} + \sigma_{ix}\sigma_{jx}\sigma_{kz}\sigma_{lz}]
 \\ \nonumber -\frac{1}{2} \sin^2 \theta_A\sin (2\theta_B) [\sigma_{ix}\sigma_{jz}\sigma_{kx}\sigma_{lx} + \sigma_{ix}\sigma_{jx}\sigma_{kx}\sigma_{lz}] + \cos^2 \theta_A \sin^2\theta_B \sigma_{iz}\sigma_{jx}\sigma_{kz}\sigma_{kz}\sigma_{lx} \\ \nonumber - \frac{1}{2} \sin (2\theta_A)\sin^2 \theta_B [\sigma_{iz}\sigma_{jx}\sigma_{kx}\sigma_{lx} + \sigma_{ix}\sigma_{jx}\sigma_{kz}\sigma_{lx}] + \sin^2 \theta_A \sin^2\theta_B \sigma_{ix} \sigma_{jx}\sigma_{kx} \sigma_{lx}  \biggr\}.
\end{eqnarray}

\noindent The anisotropic coupling constants $J_x = -J \cos \theta_A \cos \theta_B + \Delta \sin \theta_A \sin \theta_B$, $J_y = -J$, $J_z = -J \sin \theta_A \sin \theta_B + \Delta cos \theta_A \cos \theta_B$, $J_{zx} = -J \sin \theta_A \cos \theta_B -\Delta \cos \theta_A \sin \theta_B$, and $J_{xz} = -J \cos \theta_A \sin \theta_B - \Delta \sin \theta_A \cos \theta_B$. Notice that we have divided the two-dimensional lattice into even and odd sublattices, denoted by $A$ and $B$ such that $\theta_i = \theta_A$ and $\theta_j = \theta_B$ for $i \in A$ and $ j \in B$, respectively.

Second, we introduce Holstein-Primakoff bosonic operators $a_i$ and $b_j$ for sublattices $i \in A$ and $ j \in B$ via Holstein-Primakoff transformation,

\begin{eqnarray}
\sigma_{iz} = \sigma - a_i^\dagger a_i \enspace \hbox{and} \enspace
\sigma_{i}^{+} = \sqrt{1-\frac{n_i}{2\sigma}} a_i,
\end{eqnarray}

\noindent for the sublattice $A$ and,

\begin{eqnarray}
\sigma_{jz} = b_j^\dagger b_j - \sigma  \enspace \hbox{and} \enspace
\sigma_{j}^{+} = b^\dagger_j\sqrt{1-\frac{n_j}{2\sigma}},
\end{eqnarray}

\noindent for the sublattice $B$, where $n_i = a_i^\dagger a_i$ and $n_j = b_j^\dagger b_j$ are the boson number operators and $\sigma_{l}^{+} = \sigma_x + i \sigma_y \equiv (\sigma_{l}^{-})^\dagger$ is the spin raising operator.

Third, we express the spin Hamiltonian in terms of HP bosons and keep only the terms up to the quadratic order in bosonic operators. Finally, by introducing the Fourier transformation of the bosonic operators,

\begin{eqnarray}
a_k = \biggr ( \frac{2}{N} \biggr)^{1/2} \sum_l e^{i\vec{k} \cdot \vec{r}_l} a_l \\ \nonumber
b_k = \biggr ( \frac{2}{N} \biggr)^{1/2} \sum_m e^{-i\vec{k} \cdot \vec{r}_m} b_m,
\end{eqnarray}

\noindent where $N$ is the total number of lattice sites, we write the linear spin-wave Hamiltonian as $H = H_0 + H_1 + H_2$. The zeroth order that describes the classical energy is given by,

\begin{eqnarray}
H_0 = -\frac{s^2Nz}{2}[-J \sin \theta_A \sin \theta_B + \Delta \cos \theta_A \cos \theta_B] + \frac{s^4Nz}{4} J_q \cos^2 \theta_A \cos^2 \theta_B,
\end{eqnarray}

\noindent where $z = 4$ is the number of nearest neighbors and $s =1/2$ is the effective spin. By minimizing the classical energy with respect to $\theta_A$ and $\theta_B$, we find two relations,

\begin{eqnarray}
2 J_{xz} + J_qS^2 \sin (2\theta_A) \cos^2 \theta_B =0 \\ \nonumber
2J_{zx} + J_qS^2 \sin (2\theta_B) \cos^2 \theta_A = 0.
\end{eqnarray}

\begin{figure}
\includegraphics[width=\columnwidth]{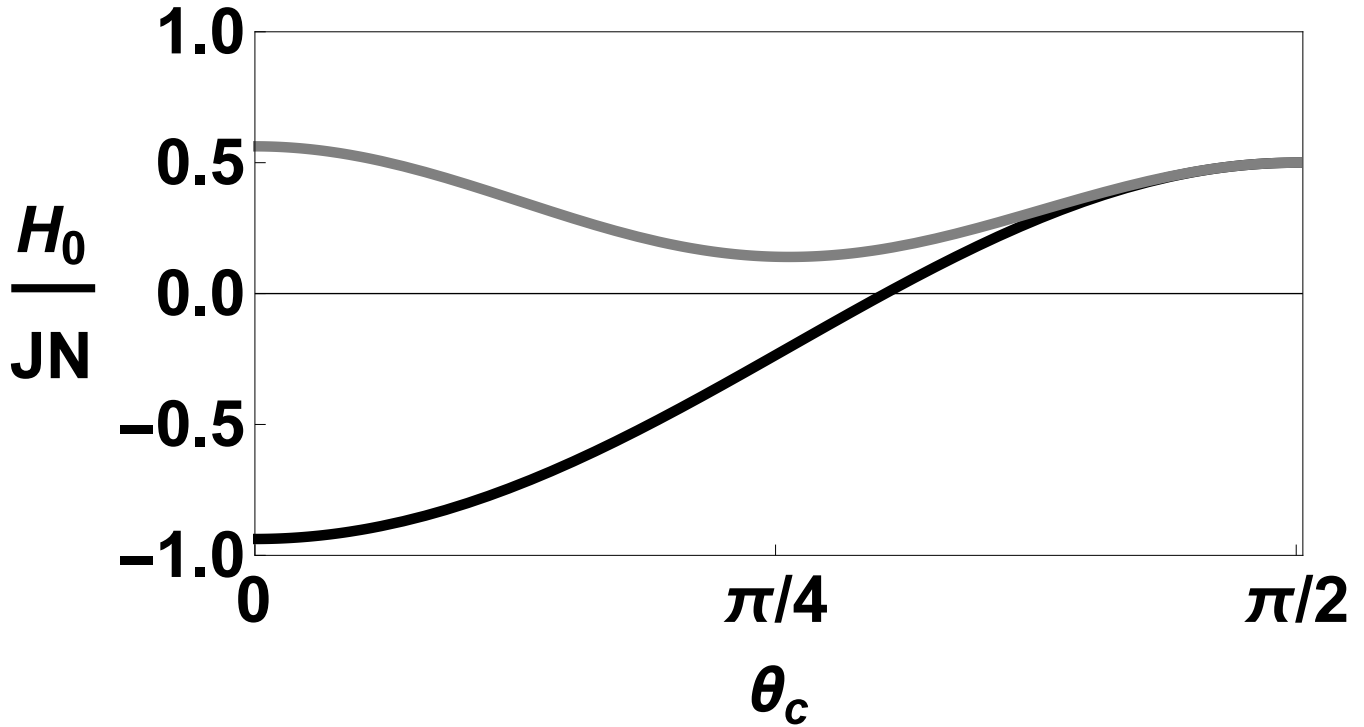}
\caption{Classical energy derived from linear spin wave theory at $\Delta = 2J$. While the black solid line is for $J_q = J$ and the gray solid line is for $J_q = 25 J$. Notice the minimum energy at canted angle $\theta_c = 0.805$ radians for $J_q = 25 J$, where as the energy is minimum at $\theta_c = 0$ in the AFM state at $J_q = J$.}\label{CE}
\end{figure}

\noindent By solving these equations simultaneously, we find two solutions $\theta_A = \theta_B = 0$ or $\theta_A = \theta_B = \cos^{-1}[\sqrt{\frac {J +\Delta}{J_qs^2}}]$, representing the AFM state and C-AFM state respectively. This indicates that the spin system undergoes a second order phase transition from an AFM state to a C-AFM state as one increases $J_q$. The phase boundary between these two states is given by the condition $J + \Delta = J_qs^2$ and the spin canted angle with respect to AFM spin orientation is given by $\theta_c = \cos^{-1}[\sqrt{\frac {J +\Delta}{J_qs^2}}]$. As a demonstration, we plot the classical energy per site as a function of canted angle for two different sets of representative parameters in FIG. \ref{CE}. As can be seen, the classical energy shows a local minimum at zero canted angle for smaller values of $J_q$, where it shows a local minimum at a finite value of canted angle ($\theta_c$) for larger values of $J_q$.

The first order terms that are linear in creation and annihilation bosonic operators are given by,

\begin{eqnarray}
H_1 = \frac{J_{zx} S^{3/2}}{N^{1/2}}\sum_{ij} \sum_k e^{-ik r_j} (b_k + b^\dagger_{-k}) - \frac{J_{xz} S^{3/2}}{N^{1/2}}\sum_{ij} \sum_k e^{-ik r_i} (a_{-k} + a^\dagger_{k}) \\ \nonumber
+\frac{J_{q} S^{7/2}}{N^{1/2}}\cos^2 \theta_A \sin (2\theta_B) \sum_{ijlm} \sum_k e^{-ik r_j} (b_k + b^\dagger_{-k})\\ \nonumber
-\frac{J_{q} S^{7/2}}{N^{1/2}}\cos^2 \theta_B \sin (2\theta_A) \sum_{ijlm} \sum_k e^{-ik r_i} (a_{-k} + a^\dagger_{k}).
\end{eqnarray}

\noindent As the spin system is in either AFM or C-AFM ground states with the angles that minimize the classical energy given by Eq. (18), all the first order terms given in $H_1$ vanish.

The second order terms of the spin wave Hamiltonian is given by,

\begin{eqnarray}
H_2 = \sum_k \biggr[J_0 (a_k^\dagger a_k + b^\dagger_kb_k) + J_1 (a_ka_{-k} + a^\dagger_k a^\dagger_{-k} + b_k b_{-k} + b^\dagger_k b^\dagger_{-k}) + J_{-} (a_kb^\dagger_{-k} + a_K^\dagger b_{-k}) + J_{+} (a_kb_k + a^\dagger_k b^\dagger_k)   \biggr]
\end{eqnarray}

\noindent where the momentum dependent coefficients in each term are, $J_0 = sJ_zz + zs^3J_q \cos^2 \theta_B (2\cos^2 \theta_A  + \gamma_k \sin^2 \theta_A)$, $J_1 = zs^3J_q \gamma_k \sin^2 \theta_A \cos^2 \theta_B$, $J_{-} = [J_xs - J_ys - s^3J_q \sin (2\theta_A) \sin (2 \theta_B)]z\gamma_k/2$, and $J_{+} = [J_xs + J_ys - s^3J_q \sin (2\theta_A) \sin (2 \theta_B)]z\gamma_k/2$. Here the structure factor $\gamma_k = 1/z \sum_\delta e^{i\vec{k} \cdot \vec{\delta}}$ is introduced with the nearest neighbor vector $\vec{\delta} = d \hat{x} + d \hat{y}$, where $d$ is the lattice constant. Finally, using the equation of motion for bosonic operators $e_k = \{a_k, b_{-k}, a_k^\dagger, b_{-k}^\dagger\}$,

\begin{eqnarray}
i\frac{de_k}{dt} = -[H_2, e_k] = \lambda_{ek} e_k,
\end{eqnarray}

\noindent we find two distinct eigenvalues $\lambda_{ak}$ and $\lambda_{bk}$, where $[X, Y]$ stands the commutator for operators $X$ and $Y$. We then determine the linear spin wave frequency $\omega_k = \sqrt{\lambda_{ak} \lambda_{bk}}$,

\begin{eqnarray}
\omega_k = \sqrt{J_0^2-(J_{+}^2 - J_{-}^2)}.
\end{eqnarray}

\begin{figure*}
\centering
\subfloat[The magnon dispersion $\omega_k$ for $\Delta = 4J$ and $J_k = J$. For these parameters the spin system is in AFM state.]{\includegraphics[width=0.5\columnwidth]{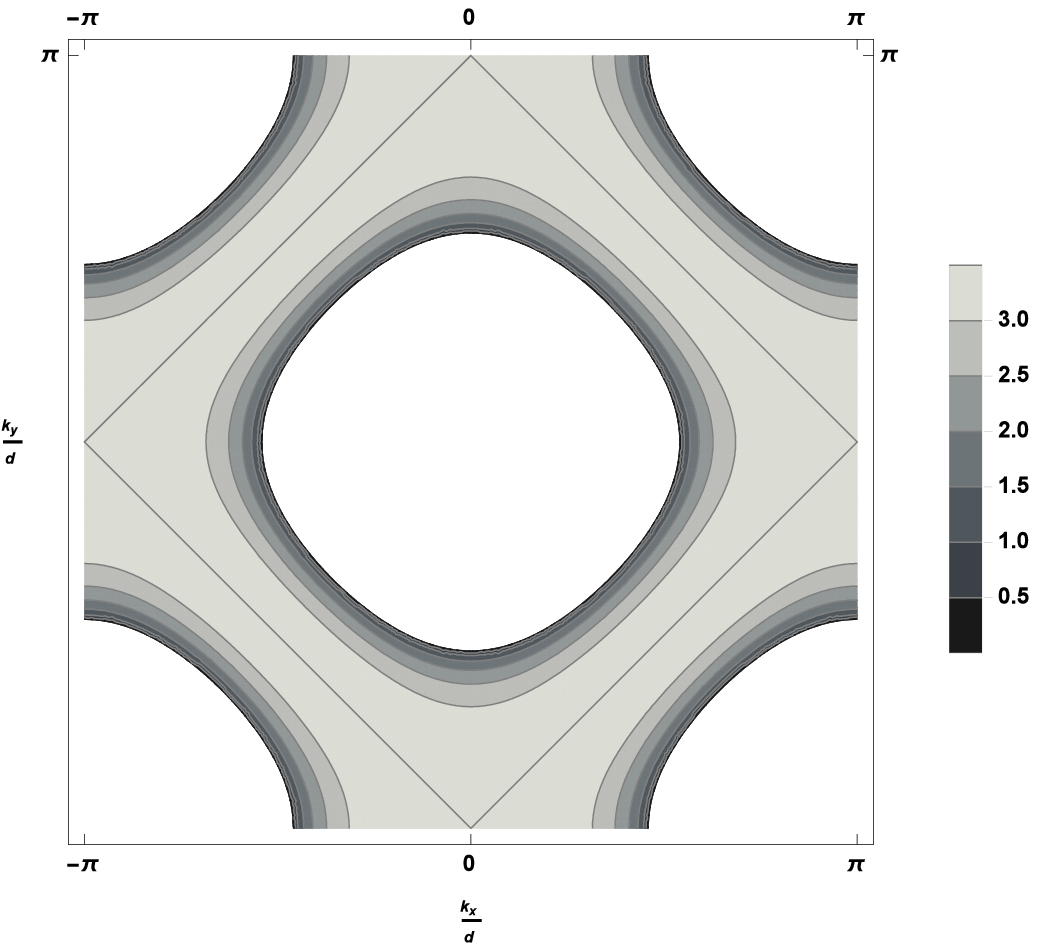}}
\subfloat[The magnon dispersion $\omega_k$ for $\Delta = 5J$ and $J_k = 25 J$. For these parameters the spin system is in C-AFM state with canted angle $\theta_c = 0.201$.]{\includegraphics[width=0.5\columnwidth]{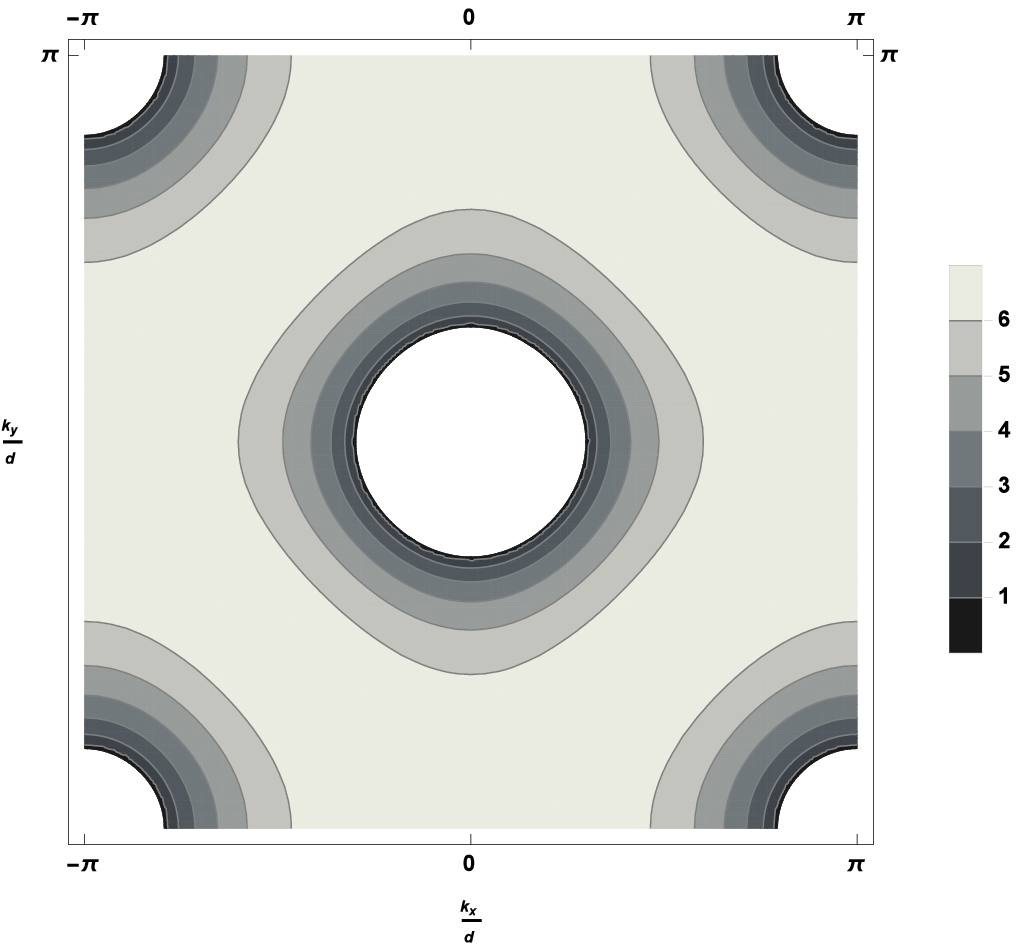}}
\caption{The magnon dispersion $\omega_k$ for AFM and C-AFM states.} \label{om}
\end{figure*}

\noindent This magnon dispersion for the AFM state and the C-AFM state is plotted in FIG.~\ref{om}. We use two sets of representative parameters to show the key features in the dispersion relevant to each of these phases. Notice that for some parameters, the magnon dispersion can be complex. This complex spin wave frequency indicates the instability of the ordered phase and represents the paramagnetic phase. This is not surprising as our spin wave theory assumes the long range magnetic order.

\section{V. CONNECTIONS TO EXPERIMENTS}

Here we considered a two-dimensional optical lattice with fully occupied lower energy $s$-orbital atoms and partially occupied higher energy $p$-orbital atoms. Such a system can be realized by loading single hyperfine state fermionic atoms, such as $^{40}$K or $^{6}$Li. After the $s$-orbital is fully occupied, partial $p$-orbital occupation is attainable at a higher density filling factors. As the anisotropic nature of the $p$-orbitals, the Bloch band formed by the $p_z$ orbital has a higher energy in two dimensions, thus the $p_x$ and $p_y$ orbitals are degenerate. The strong attractive interactions between spinless fermions can be achieved by $p$-wave Feshbach resonance.

The paramagnetic (PM), antiferromagnetic (AFM), and the canted-AFM phases elevated by the mapped spin Hamiltonian given in Eq.~(4) correspond to different particle occupation patterns relevant to the original particle Hamiltonian given in Eq.~(1). The PM phase represents the random single particle occupation of $p_x$ or $p_y$ orbitals at each site. The AFM phase represents the alternative empty and double occupation, $(0,2,0,2,...)$ pattern. Meantime, the C-AFM phase represents the random empty and double particle occupation, $(0,2,2,0,0,..)$ pattern. The canted angle measures the randomness and it depends on the microscopic parameters in the Hamiltonian. In both AFM and C-AFM states, every lattice sites are either empty or doubly occupied. Therefore, the predicted magnetic phases of the effective spin Hamiltonian can be directly detected by the density pattern in the lattice. These density patterns can be directly probed by \emph{in situ} imaging techniques~\cite{e1}. Alternatively, one can use single-site detecting techniques to map these density patterns experimentally~\cite{e2, e3}. At low temperatures, the two Fermi atoms in different orbitals at each double occupation sites can bind together to form a boson. As one lower the temperature of the optical lattice, these bosons can condensate and show normal-superfluid phase transition. The details of this phase transition is beyond the scope of this paper and we leave further details to a separate study.

\section{VI. CONCLUSIONS}

We have studied the phase diagram of doubly degenerated $p$-orbital attractive fermions on a square lattice. We considered the strongly interacting limit and the half filling density limit, and derived an effective spin Hamiltonian using a perturbation theory. We find that the fourth order square plaquette interaction term generated from the anisotropic tunneling process is significant at these limits. We used a variational mean field approach and a linear spin-wave approximation to map out the phase diagram of the effective spin Hamiltonian and find three distinct magnetic phases. Further, we discussed the experimental connections to these phases and argued that the resulting phases can be detected by using currently available experimental techniques in cold gas experiments.

\end{document}